\documentstyle[prl,aps,epsf]{revtex}\def\narrowtext{}\tighten\twocolumn
\input epsf.sty
\begin{document}
\draft

\title{
Vortex State of
Tl$_2$Ba$_2$CuO$_{6+\delta}$ via $^{205}$Tl NMR at 2 Tesla
}
\author{
Y. Itoh,$^{1}$ C. Michioka,$^{1}$ K. Yoshimura,$^{1}$ A. Hayashi,$^{2}$ and Y. Ueda,$^{2}$
}

\address{
$^1$Department of Chemistry, Graduate School of Science,\\
Kyoto University, Kyoto 606-8502, Japan \\
$^2$Institute for Solid State Physics, 
University of Tokyo, 5-1-15 Kashiwanoha,\\
Kashiwa, Chiba 277-8581, Japan\\
}

\date{\today}%
\maketitle %

\begin{abstract} 
We report a $^{205}$Tl NMR study of vortex state 
for an aligned polycrystalline sample 
of an overdoped high-$T_c$ superconductor
Tl$_2$Ba$_2$CuO$_{6+\delta}$ ($T_{c}\sim$85 K) 
with magnetic field 2 T along the $c$ axis.  
We observed an imperfect vortex lattice, so-called Bragg glass at $T$=5 K, 
coexistence of vortex solid with liquid between 10 and 60 K,
and vortex melting between 65 and 85 K.
No evidence for local antiferromagnetic ordering at vortex cores 
was found for our sample.   
\end{abstract}

\narrowtext

\section{Introduction}
\label{sec:intro} 

Coexistence of magnetism with the mixed state of type-II superconductors
has attracted great interests ~\cite{SCZhang}.  
Nearly antiferromagnetic spin fluctuation close to
the quantum critical point has been a key to
understand the magnetic correlation of high-$T_c$ superconductors.  
Such an underlying magnetic correlation may affect
the local electronic structure at vortex cores.

Tl$_2$Ba$_2$CuO$_{6+\delta}$ (Tl2201) is an overdoped high-$T_c$ superconductor. 
The optimal $T_c$ is about 85 K for a reduced oxygen content.
The early studies by Tl NMR with 4.26 T ~\cite{Mehring} 
and Cu NMR with 7 T ~\cite{AGoto} for Tl2201
reported an obvious Redfiled pattern due to the vortex lattice. 
Although the magnetic ordering
in the mixed state is not expected for Tl2201, 
because it is a typical overdoped electronic system 
away from antiferromagnetic instability \cite{KambePRB},
the recent Tl NMR study with 2.1 T shows 
coexistence of local antiferromagnetic ordering
at the vortex cores ~\cite{Kaku}. 
The specific characteristics of Tl2201 are 
cation disorder on the Tl site ~\cite{Shima} and 
magnetic field dependence of the elementary excitations ~\cite{KambePRL,OV,KambePRB2}.    
Disorder makes the vortex phase rich ~\cite{Koba}.    
The effect of these characteristics on the antiferromagnetic vortex core
is poorly understood. 
 
In this paper, we studied the vortex state 
of an overdoped high-$T_{c}$ superconductor
Tl2201 ($T_{c}\sim$85 K) 
with magnetic field of 2 T along the $c$ axis
using a $^{205}$Tl NMR spin-echo technique. 
We observed a Redfield pattern of the $^{205}$Tl NMR spectrum at 5 K 
due to the vortex lattice state 
without static antiferromagnetic order at the vortex cores. 

\section{Experiments}
\label{sec:expt}

The powder sample in the present study is the same one as in the previous
study ~\cite{KambePRB}, which was mixed with Stycast 1266 epoxy 
and magnetically aligned along the $c$ axis.    

A phase-coherent-type pulsed spectrometer was utilized to perform the $^{205}$Tl 
NMR (spin $I$=1/2) experiments.   
All the measurements were done with magnetic field cooling. 
The NMR frequency spectra with quadrature detection were obtained by
integration of the nuclear spin-echoes with frequency $\nu$.
The width of the first exciting $\pi$/2-pulse $t_{w}$ was 4
$\mu$s (the excited frequency region $\nu_1\sim$63 kHz).  
Because of no fine structure but moderately broad NMR spectra, 
high spectral resolution ~\cite{Clark} was not required in the present measurements. 
   
The nuclear spin-lattice relaxation curves
$p(t)\equiv 1-M(t)/M(\infty)$ (recovery curves) were measured 
by an inversion recovery technique,
as functions of a time $t$ after an inversion pulse, 
where the nuclear spin-echo amplitude $M(t)$, $M(\infty)[\equiv M(10T_1)]$ and $t$ 
were recorded. 
In contrast to the previous study ~\cite{Kaku}, 
the comb pulse train was not needed. 
The measurement of $M(\infty)$ was necessary
to determine the function form of $p(t)$.  

\section{Results and discussion}
\subsection{NMR spectrum}

 Figure 1 shows $^{205}$Tl NMR spectra 
with magnetic field of $H_{0}$=2 T along the $c$ axis
at $T$=85 and 5 K.  
The bare resonance line $\nu_{0}$ 
given by $^{205}\gamma_nH_0$ ($^{205}\gamma_n$=24.567 MHz/T)
is denoted by the dash line. 
Above $T_{c}$, the full width at half maximum of the spectrum 
is about 50 kHz, being nearly the same as the reported one
~\cite{Kaku}. 
The linewidth can be considered to result from 
a staggered spin susceptibility (the second moment)
and from a Knight shift distribution (the first moment).
The enhancement of the static staggered spin susceptibility along the $c$ axis
is estimated from 
the $^{63}$Cu nuclear spin-spin relaxation rate 1/$T_{2G}$ ~\cite{Itoh}. 

The $^{205}$Tl Knight shift is understood by
the sum of the spin shift $K_{c, spin}$ 
due to the Cu-to-Tl supertransferred hyperfine coupling
and the temperature-independent $K_{c, res}$
due to residual spin density polarization ~\cite{KambePRB2}. 
Below $T_{c}$, the spectrum peak is largely shifted into 
the negative shift side (low frequency region), 
while the tail is still in
the positive shift side (high frequency region). 
Since for a singlet $d_{x^2-y^2}$-wave superconductor, 
local density of states of electrons is suppressed away from vortex cores but
recovers at the cores ~\cite{TIM}, 
then we expect that a Redfield pattern of 
the spin shift $^{sc}K_{c, spin}(\nu)$ 
due to the electron density of states 
affects the NMR spectrum.  
However, the observed negative shift indicates that a Redfield pattern
of a superconducting diamagnetic shift $^{sc}K_{c, dia}(\nu)$
is predominant.  

The $T$=5 K NMR spectrum resembles the field distribution 
due to a slightly distorted vortex lattice ~\cite{Riseman,Brandt}
but not a Gaussian due to amorphous vortex or vortex glass. 
We could not deduce from the NMR spectrum
which the vortex lattice is, triangular or square.  

The peak frequency can be assigned to the saddle point (S) 
in the field distribution of vortices,
the highest frequency edge to the vortex core (V), and 
the lowest frequency edge to the minimum field position (C) 
mostly away from the vortex cores ~\cite{TIM,Red}. 
Site-selective NMR is possible for our sample. 

In Fig. 1, the Knight shift at the vortex core V at 5 K
is nearly the same as or larger than the normal shift at 85 K. 
An external magnetic field
is squeezed at the vortex core. 
We notice that the squeezed magnetic flux density at the core
yields a higher magnetic field than an applied external field,
so that the shift at the core-site V can be larger than the normal shift.
Using the additional positive vortex core shift 
$^{sc}K_{c, field}$(V) due to the enhanced field and
the negative shift 
$^{sc}K_{c, dia}$(C)$<$0 due to the diamagnetic field,
we obtain 

\begin{equation} 
\nu(\mathrm{V})=
\nu_{0}[1+^{sc}K_{c, spin}(\mathrm{V})][1+^{sc}K_{c,
field}(\mathrm{V})]   
\label{e.core}
\end{equation}
with $^{sc}K_{c, spin}$(V)$\geq$
$K_{c, spin}$($T_{c}$) ~\cite{TIM}, 
while 
\begin{equation} 
\nu(\mathrm{C})=\nu_{0}[1+^{sc}K_{c, dia}(\mathrm{C})].
\label{e.mini}
\end{equation}
These are illustrated in Fig. 2. 
A part of the NMR
signals at $\nu(\mathrm{V})\geq$49.18 MHz is
due to the combined effect of magnetic flux squeezing and of the finite positive spin
shift at the vortex core. 
To our knowledge, no one has ever pointed out the importance of this effect.   

No signal above 49.4 MHz at 5 K, that is,  
no evidence for the local antiferromagnetic ordering
effect at the vortex cores was found for our sample.
The difference in imperfection or in the crystal symmetry ~\cite{Shima}
may affect the local order. 

\subsection{Nuclear spin-lattice relaxation}

Figure 3(a) shows frequency distribution of
the recovery curves at 5 K.  
The recovery curve is close to an exponential function 
at the peak frequency (S)
but nonexponential at lower and higher frequency sides. 
These results remind us of the in-plane impurity-substitution effect
~\cite{Itoh2}.  
The nuclear moments close to an
impurity show nonexponential relaxation,
because of random impurity distribution.  

Figure 3(b) shows temperature dependence of the recovery curve 
at $\nu$=49.18-49.20 MHz. One can see nonexponential curves below 30 K
and a nearly exponential one at 40 K. 

The solid curves in Fig. 3 are the least-squares fitting results by 
\begin{equation} 
p(t)=p(0)\mathrm{exp}[- \frac{t}{T_1}-\sqrt{\frac{t}{\tau_1}}], 
\label{e.stre}
\end{equation}
where $p(0)$, $T_1$ and $\tau_1$ are the fitting parameters. 
Equation (\ref{e.stre}) is derived from  
an impurity-induced NMR relaxation theory ~\cite{Mac}. 
$T_1$ is a nuclear spin-lattice relaxation time
due to quasi-particle scattering or homogeneous relaxation process, 
whereas $\tau_1$ is an induced 
relaxation time due to random distribution of impurity relaxation centers. 
Since there are no literal localized moments, 
the finite $\tau_1$ may result from  
an extended staggered spin fluctuation around the vortex core ~\cite{Tsuchi},
Friedel oscillation around the core ~\cite{MM},
or thermal vortex motion ~\cite{Furman}.

For breakdown of the site-selective NMR,
the recovery curve of the nuclear magnetization
is nonexponential in the mixed state ~\cite{Furman}. 
In the case that
the magnetic penetration depth $\lambda_c$ is too long, 
a sharp NMR line with hierarchical $T_1$ components is observed and then 
the site-selective NMR does not work.   
The second moment of the NMR spectrum in the vortex lattice, 
a guide to broadening, is given by 
\begin{equation} 
\sqrt{<\Delta H_c^2>}\propto 1/\lambda_c^2, 
\label{e.fwhm}
\end{equation}
where $\Delta H_c$ is the field distribution around the peak ~\cite{Pincus}. 
For deeply underdoped and heavily overdoped superconductors,
the long $\lambda_c$'s are estimated ~\cite{Nie}.  
For our sample,
the observed Redfield pattern indicates a short $\lambda_c$. 
Thus, we suppose one-to-one correspondence of the site to the frequency shift.  

 Let us briefly discuss the reason why the relaxation curve
is nonexponential in spite of site selection in NMR.
Figure 4(a) illustrates a perfect square lattice of vortices
on a CuO$_2$ plane for convenience. 
The recovery curve at any nuclear site in Fig. 4(a)
must be an exponential function,
even if local moments or local staggered moments are induced at the vortex cores.
For example, the vortex configurations around all the B sites 
in Fig. 4(a) are the same, so that all the recovery curves 
of the B-site nuclei would become the same.
However, we observed the nonexponential recovery curves at the selected sites.  
Thus, the vortex lattice is not perfect.
One may call it Bragg glass ~\cite{BG}.  
Figure 4(b) illustrates an imperfect vortex lattice with weak disorder. 
The vortex configuration around the B site in Fig. 4(b) 
is not unique, so that the assemble average on the vortex distribution
leads to nonexponential relaxation.  

\subsection{Coexistence of vortex solid with liquid}

Figure 5(a) shows $^{205}$Tl NMR spectra with magnetic field cooling. 
The narrow NMR lines between $T_c$=85 K and $T$=65 K indicate
motional narrowing effect, that is,   
vortex melting. 
The NMR spectra between $T$=60 and 10 K are composed of two peaks 
with a negative and a positive Knight shifts. 
These peaks indicate coexistence of the vortex solid with liquid,
that is, partial melting. 
At $T$=5 K, the Redfield pattern indicates the vortex lattice. 
The similar, but narrow, coexistence region was first reported 
for YBa$_2$Cu$_3$O$_7$ ~\cite{Reyes1,Reyes2}. 
The detailed measurement of vortex phase 
is in progress for our Tl2201 sample.   

Figures 5(b) and 5(c) show 1/$\tau_1$ and 1/$T_1$ 
as functions of temperature
at C, S, and L on the solid lines in Fig. 5(a).  
In Fig. 5(b), the peak behavior of 1/$\tau_1$ at L 
is observed at $T_f$=20 K.  
Since the signal L comes from the vortex liquid in the coexistence region,
the peak behavior indicates
slowing down of vortex motion toward freezing. 
This is not due to
the local antiferromagnetic ordering around vortex cores. 
$T_f$ is close to the irreversibility line ~\cite{Berg}.
The temperature dependence of 1/$T_1$ at C and S in Fig. 5(c) 
can be understood by the spatial dependence of 
local density of states of electrons ~\cite{TIM}. 

More than ten years have passed
since our sample was synthesized. 
The vortex phase is sensitive to 
crystalline imperfection. 
The aging effect, e.g. oxygen redistribution, 
might realize the solid-liquid coexistence.
The effect, however, does not cause $T_c$ distribution
in our sample. 
We exclude a possibility that
the signal L is a lower-$T_c$ part of the sample. 
The magnetic susceptibility does not exhibit 
$T_c$ distribution.  
The Cu nuclear quadrupole resonance spectrum at $T$=4.2 K,
which is sensitive to the oxygen distribution,
is still similar to the previous report ~\cite{KambeC}.  
The temperature dependence of 1/$\tau_1$ 
of the signal L is not 
conventional behavior of superconducting onset.   

\section{Conclusion}
\label{sec:conclusion}
To conclude, we observed a vortex lattice with weak disorder 
for our sample Tl2201 ($T_{c}\sim$85 K) with 2 T at 5 K,
coexistence of vortex solid with liquid ($T$=10-60 K),
and vortex liquid ($T$=65-85 K), but not   
the local antiferromagnetic ordering at the vortex cores.

\acknowledgments
We thank A. Goto for fruitful discussion,
and M. Kato and T. Waki for experimental supports.   
This was supported by a Grant-in-Aid
on priority area 
"Novel Quantum Phenomena in Transition Metal Oxides," 
from Ministry of Education, Science,
Sports and Culture (12046241).

\begin{figure}
\epsfxsize=2.9in
\epsfbox{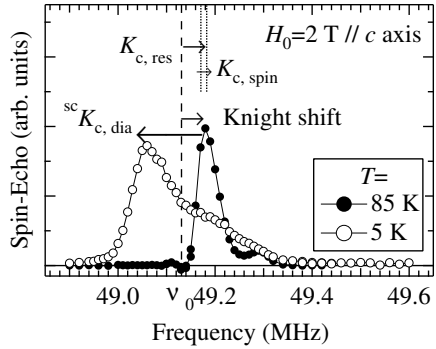}
\vspace{0.0cm}
\caption{
$^{205}$Tl NMR spectra with field cooling 
(2 T along the $c$ axis)
at $T$=85 and 5 K.  
$\nu_0$ is given by $^{205}\gamma_nH_0$ ($^{205}\gamma_n$=24.567 MHz/T). 
The small signal around 49.3 MHz at 85 K 
is probably due to misalignment of some of the grains.  
}
\label{TLNMR}
\end{figure}
 
\begin{figure}
\epsfxsize=2.8in
\epsfbox{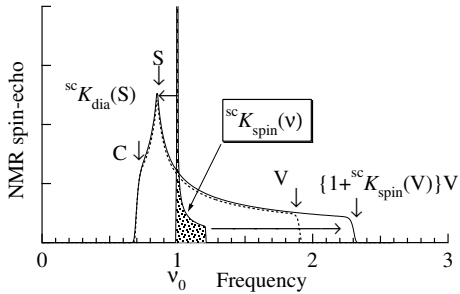}
\vspace{0.0cm}
\caption{
Illustration of NMR spectrum in a vortex state.
Superconducting diamagnetic shifts (dash curve),
$d$-wave spin shifts (shaded region), and
their combination (solid curve).      
}
\label{TLNMR}
\end{figure}
 
\begin{figure}
\epsfxsize=3.5in
\epsfbox{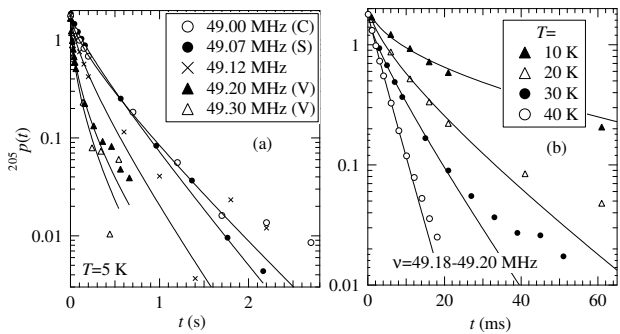}
\vspace{0.0cm}
\caption{
Recovery curves of $^{205}$Tl nuclear spin-echoes
against frequency at $T$=5 K (a) and 
those against temperature at $\nu$=49.18-49.20 MHz (b). 
The solid curves are the least-squares fits by Eq. (\ref{e.stre}).  
}
\label{Recovery}
\end{figure}
 
\begin{figure}
\epsfxsize=3.3in
\epsfbox{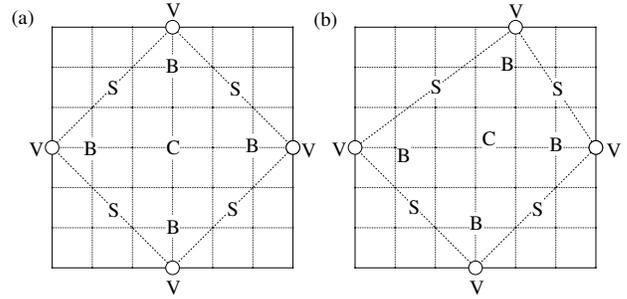}
\vspace{0.0cm}
\caption{
Illustration of vortex lattices on a CuO$_2$ plane, 
(a) a perfect square lattice and (b) an imperfect square lattice.
Open circles are the vortex cores. 
The site notation conforms to Ref. [13].
}
\label{Vortex}
\end{figure} 

\begin{figure}
\epsfxsize=3.4in
\epsfbox{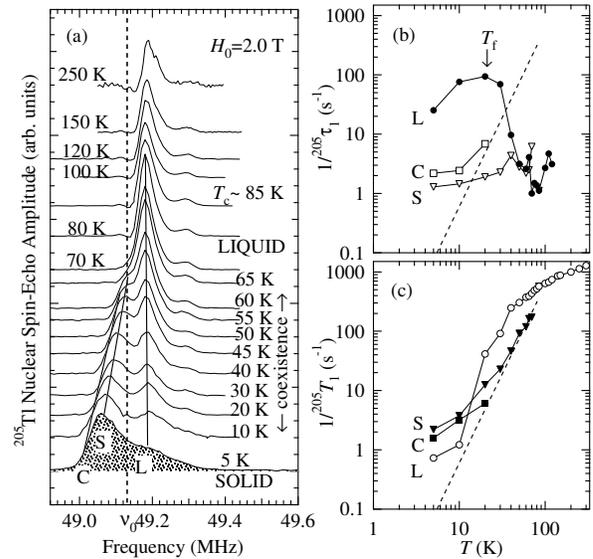}
\vspace{0.0cm}
\caption{
(a) $^{205}$Tl NMR frequency spectra, 
(b) 1/$\tau_1$, and (c) 1/$T_1$ with magnetic field cooling.  
Vortex melting (LIQUID), coexistence of vortex solid with liquid (coexistence),
and vortex lattice (SOLID) are denoted in (a). 
No appreciable frequency distribution of the $^{205}$Tl nuclear spin-echo
decay curve was found at any temperatures. 
The $^{205}$Tl nuclear spin-lattice relaxation rates
were measured at C, S, and L on the solid lines in (a). 
The dash lines in (b) and (c) are $T^3$ due to
a superconducting gap with line nodes. 
}
\label{Vortex}
\end{figure}

\end{document}